\documentclass{appolb}
\usepackage{epsfig}

\newcommand{\aindex}{{\scriptscriptstyle A}}
\newcommand{\lindex}{{\scriptscriptstyle L}}
\newcommand{\tindex}{{\scriptscriptstyle T}}
\newcommand{\tr}{\mbox{Tr}\,}

\newcommand{\derslash}{\rlap{/} \partial}
\newcommand{\Aslash}{\rlap{/} A}
\newcommand{\Vslash}{\rlap{/} V}

\begin{document}
\title{Estimate of the Collins function in a chiral invariant approach
}
\author{\underline{A.~Bacchetta}$^1$, R.~Kundu$^2$, A.~Metz$^1$, P.J.~Mulders$^1$
\address{$^1$ Division of Physics and Astronomy, Faculty of Science, Free University \\
De Boelelaan 1081, NL-1081 HV Amsterdam, the Netherlands}
\address{$^2$ Department of Physics, RKMVC College \\
Rahara, North 24 Paraganas, India}
}
\maketitle
\vspace{-0.5cm}
\begin{abstract}
We estimate the Collins function at a 
low energy scale by calculating the 
fragmentation of a quark into a pion at the one-loop level
in the chiral invariant model of Manohar and Georgi.
We give a useful parametrization of our results and we briefly discuss 
different spin and/or azimuthal asymmetries containing the Collins function 
and measurable in semi-inclusive DIS 
and $e^+ e^-$ annihilation.
\end{abstract}

The Collins function $H_1^{\perp}$, describing the fragmentation of a 
transversely polarized quark into an unpolarized hadron \cite{collins_93},
plays an important role in studies of the nucleon spin structure.
Although $H_1^\perp$ is T-odd, it can be nonzero due to final 
state interactions.
In fact, we were able to show for the first time that a nonvanishing 
Collins function can be obtained in a field theoretical approach 
through a consistent one-loop calculation of the 
fragmentation process
\cite{bacchetta_01}.
\\
To obtain a reasonable estimate of the Collins function,
we calculate it 
in a chiral invariant approach at a low energy scale \cite{bacchetta_02}.
We use the model of Manohar and Georgi \cite{manohar_84}, which incorporates
chiral symmetry and its spontaneous breaking, two important aspects of QCD at low
energies.
\\
The Collins function enters several spin and azimuthal asymmetries in one-particle
inclusive DIS. 
Of particular interest is the transverse single spin asymmetry, where $H_1^\perp$ 
appears in connection with the transversity distribution, for which we predict 
effects of the order of $10\%$.
In principle in semi-inclusive DIS only the shape of the Collins function can be 
studied, while a purely azimuthal asymmetry in $e^+ e^-$ annihilation allows one to 
measure also its magnitude.
For this asymmetry we obtain effects of the order of $5\%$.

\section{Model calculation of the Collins function}

\begin{figure}[t]
\centering
\epsfig{figure=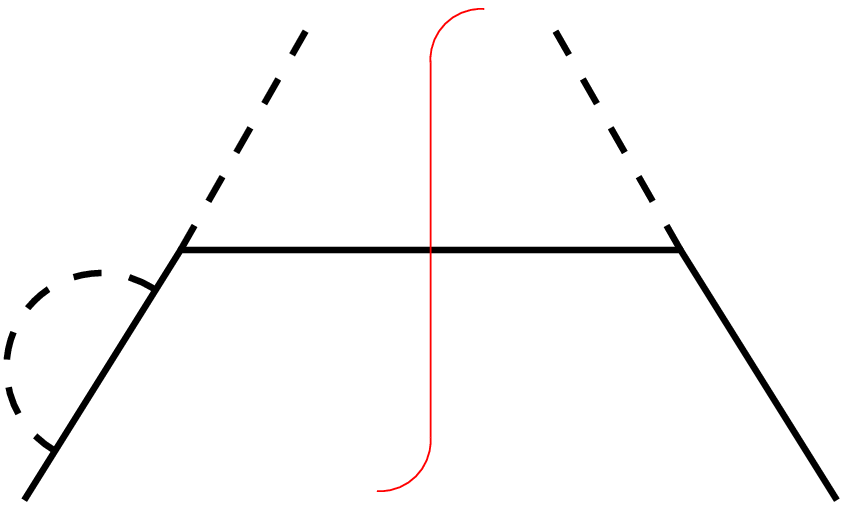,width=2.5cm}\hfil
\epsfig{figure=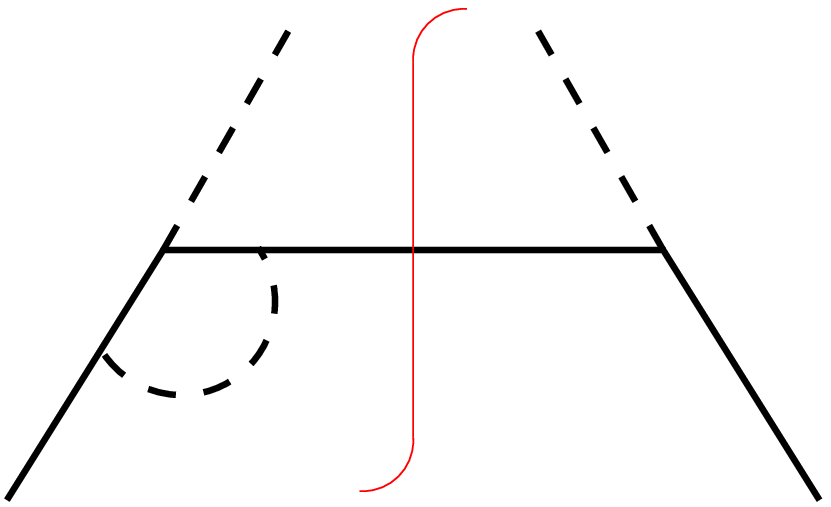,width=2.5cm}\hfil        
\epsfig{figure=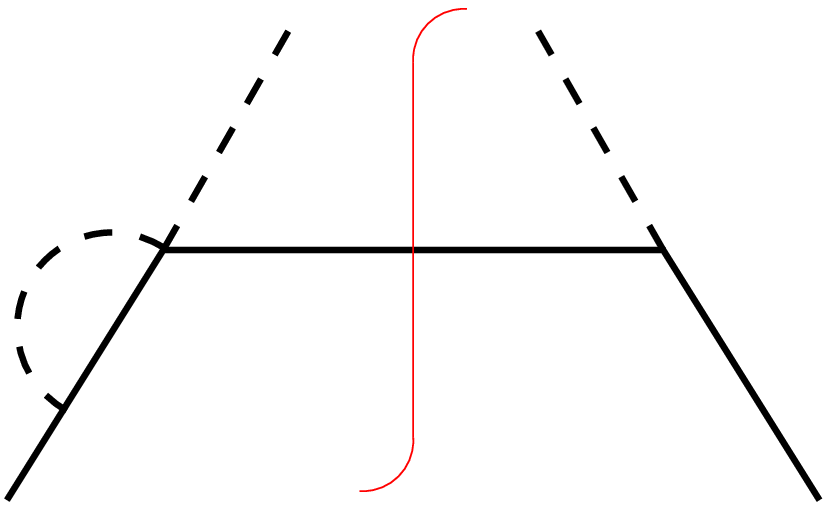,width=2.5cm}        
\caption{One-loop corrections to the fragmentation of a quark  into a pion 
 contributing to the Collins function.
 The Hermitian conjugate diagrams are not shown. \label{f:loop}}
\end{figure}

\begin{figure}[t]
\centering
\epsfig{figure=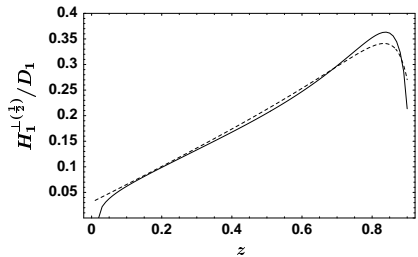,width=6cm}
\caption{Model result for the ratio $H_1^{\perp (1/2)} / D_1$ as a function of 
$z$ and the simple analytic parametrization of \protect{Eq.~(\ref{e:param})}. 
\label{f:ratio}}
\end{figure}

Considering the fragmentation process $q^{\ast}(k) \to \pi(p) X$ we define the 
Collins function, which depends on the longitudinal momentum fraction $z$ of the pion 
and the transverse momentum $k_{\tindex}$ of the quark, as \cite{mulders_96} 
\begin{equation} 
\frac{\epsilon_{\tindex}^{ij} k_{\tindex\,j}}{m_{\pi}}
 \, H_1^{\perp}(z,z^2 k^2_{\tindex}) =
 \frac{1}{4z} \int d k^+ \; \tr[ \Delta (k,p) i \sigma^{i-} \gamma_5]
 \Big |_{k^-=p^-/z} \;,  
\end{equation}
with the correlation function
\begin{equation} 
\Delta(k,p)=\sum_X\, \int
 \frac{d^4 \xi}{(2\pi)^{4}} e^{+ i k \cdot \xi}
 \langle 0| \,\psi(\xi)|\pi, X\rangle 
 \langle \pi, X| \bar{\psi}(0)|0\rangle \,.     
 \label{e:corr}
\end{equation} 
\\
To get an estimate of the Collins function at a low energy scale
we make use of the chiral invariant model of Manohar and 
Georgi \cite{manohar_84}. The model contains massive constituent quarks and
Goldstone bosons as effective degrees of freedom. The Lagrangian of the model
reads
 \begin{equation}
{\cal L} =  \bar{\psi} \, ( i \derslash + \Vslash - m 
 + g_{\aindex} \Aslash \gamma_5 ) \, \psi \,.
\label{e:lagrangian}
\end{equation}
The Manohar-Georgi model describes the valence part of the 
normal unpolarized fragmentation 
function $D_1$ fairly well \cite{bacchetta_02}.
\\
While at tree level the Collins function is zero, the situation changes if
rescattering corrections are included in $\Delta(k,p)$.
In a consistent one-loop calculation the diagrams in Fig.~\ref{f:loop}
contribute to $H_1^\perp$. Because of the presence of nonvanishing imaginary
parts in these diagrams, we obtain a nonvanishing Collins function.
\\
The $z$ behaviour of asymmetries containing $H_1^\perp$ is typically governed 
by the ratio
\begin{equation}
\frac{H_1^{\perp (1/2)}(z)}{D_1(z)} \equiv \frac{\pi}{D_1(z)} \, z^2 \, 
 \int d k_{\tindex}^2 \, \frac{|\vec{k}_{\tindex}|}{2 \, m_\pi} \, 
 H_1^\perp(z,z^2 k_{\tindex}^2) \,.
\end{equation}
In our model, it turns out that this ratio is approximately equal to
\begin{equation}
 \frac{H_1^{\perp (1/2)}(z)}{D_1(z)} \approx \frac{\langle |\vec{K}_{\tindex}|\rangle (z)}{2 z m_{\pi}}
	 \frac{H_1^{\perp}(z)}{D_1(z)} 
\approx \frac{\sqrt{ 0.9 \langle {K}_{\tindex}^2\rangle (z)}}{2 z m_{\pi}}
	 \frac{H_1^{\perp}(z)}{D_1(z)} 
\end{equation} 
For convenience of use, the result of our model can be roughly parametrized
by means of a simple analytic form, i.e.\
\begin{equation} 
\frac{H_1^{\perp (1/2)}(z)}{D_1(z)} \approx 0.316\, z + 0.0345\,\frac{1}{1-z}-
0.00359\, \frac{1}{(1 - z)^2}.
\label{e:param}
\end{equation} 
Fig.~\ref{f:ratio} shows the result of our model 
together with the parametrization we suggested. 
The ratio $H_1^{\perp (1/2)}/D_1$
is clearly increasing with increasing $z$; this feature is largely
independent of the parameter choice in our approach \cite{bacchetta_02}.

\section{Observables}
\begin{figure}[t]
\centering
\epsfig{figure=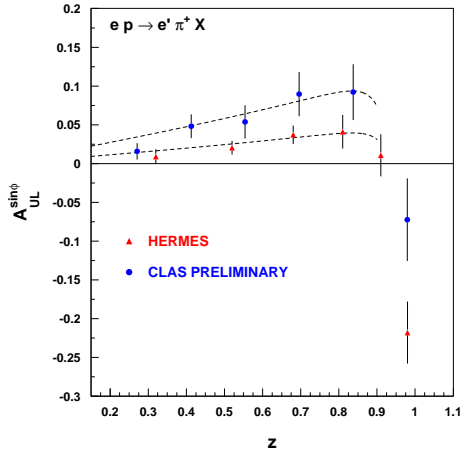,width=6cm}
\caption{Comparison between 
the results of our model and data from the HERMES and 
CLAS experiments.} 
\label{f:fit}
\end{figure}

\begin{figure}[t]
\centering
\epsfig{figure=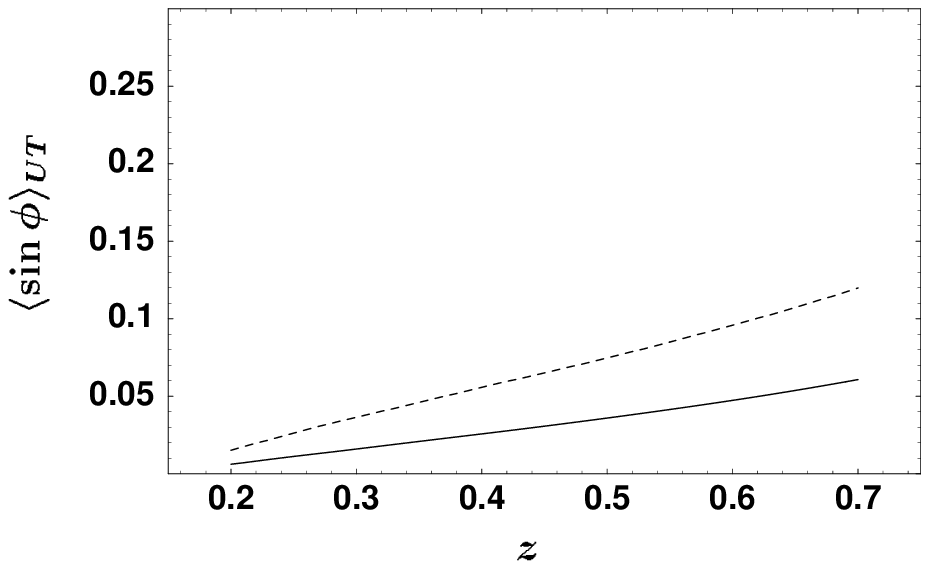,width=6cm}\hfil
\epsfig{figure=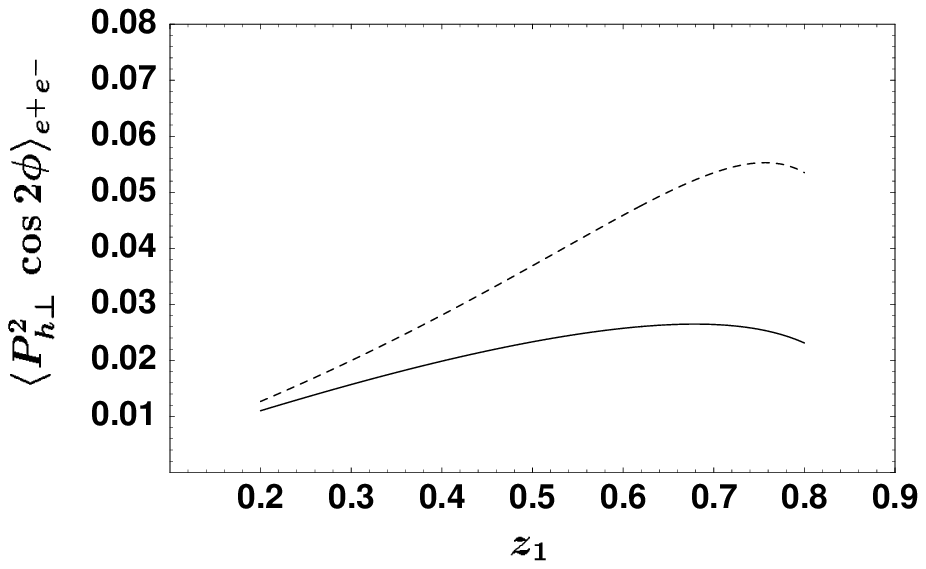,width=6cm}
\caption{Left panel: spin asymmetry $\langle\sin{\phi}\rangle_{UT}$ of 
Eq.~(\ref{e:ut}) as a function of $z$, assuming $h_1 = g_1$ (solid line), 
and assuming $h_1 = (f_1+ g_1)/2$ (dashed line).
Right panel: azimuthal asymmetry 
$\langle P_{h \perp}^2 \cos{2 \phi}\rangle_{e^+ e^-}$ of Eq.~(\ref{e:epem}), 
integrated over the angular range $0 \leq \theta \leq \pi$, and over  
the $z_2$ ranges $0.2 \leq z_2 \leq 0.8$ (solid line) and
$0.5 \leq z_2 \leq 0.8$ (dashed line). \label{f:asymm}} 
\end{figure}

In semi-inclusive DIS, three important single spin and azimuthal asymmetries
containing the Collins function exist.
Neglecting quark mass terms one has \cite{mulders_96}
\begin{eqnarray}
\langle \sin \phi_h \rangle_{UL} & \propto & 
 \frac{1}{Q} \Big[ \Big( c_1 \, h_{\lindex}(x) + c_2 \, h_1(x) \Big) 
                  H_1^{\perp (1/2)}(z) \,
                  + \, \rm{other \; terms} \Big] , \hphantom{M}
 \label{e:ul}
\\
\langle \sin \phi_h \rangle_{LU} & \propto & 
 \frac{1}{Q} \; e(x) \, H_1^{\perp (1/2)}(z) \,,
 \label{e:lu}
\\
\langle \sin (\phi_h + \phi_S) \rangle_{UT} & \propto & 
 h_1(x) \, H_1^{\perp (1/2)}(z) \,. 
 \label{e:ut}
\end{eqnarray}
where only the asymmetry numerators are written down.
$\phi_h$ ($\phi_S$) is the azimuthal angle of the produced hadron 
(target spin); the subscript $UL$, e.g., indicates an 
unpolarized beam and a longitudinally polarized target;
$h_1$ denotes the transversity distribution and $h_\lindex$,$e$ denote 
two twist-3 distributions; $c_1$ and $c_2$ are kinematical factors.
\\
From the asymmetry (\ref{e:ul}) it is difficul to draw any conclusion on the
Collins function. However, 
if we assume that the {\it other terms} in (\ref{e:ul}) are small, then the
$z$ dependence of the asymmetry should be almost entirely due to the Collins
function. In Fig.~\ref{f:fit} we compare the parametrization (\ref{e:param}) 
of our results with
HERMES data on $\langle \sin \phi_h \rangle_{UL}$ \cite{airapetian_01} and
preliminary data on the same asymmetry from CLAS \cite{avakian_02}. The
agreement of the $z$ shape is remarkable. Note that we
arbitrarily normalized our curve to take into account the unknown distribution 
functions and prefactors. 
\\
The transverse spin asymmetry in Eq.~(\ref{e:ut}) is
 one of the most 
promising observables to measure $h_1$. In fact, it is one of the ambitions of 
the HERMES Collaboration to extract the transversity distribution from this
observable. However, without any knowledge on the Collins function 
it is impossible to evaluate the magnitude of the asymmetry.
Using our model and two extreme assumptions on the transversity distribution,
we obtain values up to about $10\%$ for the asymmetry
(left panel of Fig.~\ref{f:asymm}), which therefore
should be observable. Our results
support the intention of extracting the transversity in this way.
\\
At present, data on semi-inclusive DIS allow us to extract in an assumption-free
way only the shape of $H_1^\perp$.
The azimuthal asymmetry \cite{boer_98,bacchetta_02}
\begin{equation} 
\langle P_{h \perp}^2 \cos{2 \phi} \rangle_{e^+ e^-}  
 = \frac{2 \, \sin^2 {\theta}}{1+\cos^2{\theta}} \,
\frac{H_1^{\perp (1)} (z_1)\,\bar{H}_1^{\perp (1)}(z_2)}
{\left( D_1 (z_1)\,\bar{D}_1^{(1)}(z_2)
	+ D_1^{(1)} (z_1)\,\bar{D}_1(z_2)\right)} \,,
\label{e:epem}
\end{equation}
accessible in $e^+ e^-$ annihilation into two hadrons, can provide 
information on the magnitude of $H_1^\perp$ as well.
As shown in the right panel of Fig.~\ref{f:asymm}, our result for this asymmetry 
is of the order of $5\%$. 
Possible accurate measurements of this observable at BABAR or BELLE would be very 
useful to better pin down the Collins function from the experimental side.

\end{document}